\documentclass[3p,times,twocolumn]{elsarticle}

\usepackage{ecrc}

\volume{00}

\firstpage{1}

\journalname{Nuclear Physics B Proceedings Supplement}

\runauth{P. W. Gorham}

\jid{nuphbp}

\jnltitlelogo{Nuclear Physics B Proceedings Supplement}

\usepackage{amssymb}

\usepackage[figuresright]{rotating}

\begin{document}

\begin{frontmatter}



\dochead{}

\title{Particle Astrophysics in NASA's Long Duration Balloon Program}


\author{Peter W. Gorham}

\address{Dept. of Physics and Astronomy, Univ. of Hawaii at Manoa, 2505 Correa Rd., Honolulu, HI, 96822, USA}

\begin{abstract}
A century after Victor Hess' discovery of cosmic rays, balloon flights still play a central role in
the investigation of cosmic rays over nearly their entire spectrum. We report on
the current status of NASA balloon program for particle astrophysics, with particular
emphasis on the very successful Antarctic long-duration balloon program, and 
new developments in the progress toward ultra-long duration balloons.
\end{abstract}


\end{frontmatter}


\section{Introduction}
\label{intro}

It is remarkable in this current generation of highly developed spacecraft exploration of 
the solar system that scientific ballooning still remains a powerful and effective method of
particle astrophysics investigation. More than a century after balloon flights were
first used~\cite{Hess1912} to establish the extraterrestrial nature of cosmic rays,
balloon-based investigations still play a crucial role in their elucidation.
Since the time of Viktor Hess' initial studies, a myriad of new techniques for 
the study of high energy radiation from the cosmos has been developed,
including sophisticated methods utilizing direct detection, Cherenkov, and fluorescence
radiation, observed on the ground; and particle detection in a wide range of
spacecraft. However, the unique role of cosmic ray investigations from balloons has not yet 
been superseded, for a variety of reasons: from the relatively low cost and risk for
development of new and novel instruments as balloon payloads,  the large lift
capacity of a balloon system for modest costs, and in some cases, due to
the fact that the suborbital trajectory of the balloon system is a necessary
part of the science requirements.

The term {\it particle astrophysics} was initially synonymous with cosmic ray studies,
but within the last several decades it has become increasingly evident that 
still-undiscovered fundamental particles may play a dominant role in the 
composition of the non-baryonic dark matter, which makes up of order a quarter
of the closure density of the universe. Thus the scope of particle astrophysics
has expanded to include dark matter investigations, although we cannot yet be
certain that the dark matter will yield to a particle physics solution.

In addition to the study of charged particles, high energy gamma-rays, and possible neutrons of extra-solar 
origin, neutrinos have now also emerged as another member of the particle astrophysics
panoply. The fluxes of high energy neutrinos in the TeV to EeV range are closely 
coupled to cosmic acceleration processes, and may provide ``smoking-gun'' evidence for the'
sources of both galactic and extragalactic cosmic rays. For the latter,
cosmogenic EeV neutrinos are direct byproducts of intergalactic scattering of the highest energy cosmic rays,
and have the potential to probe cosmic ray accelerators even back to the earliest epochs of
cosmic ray sources. 

\begin{figure*}[htb!]
\begin{center}
\includegraphics[width=0.8\textwidth]{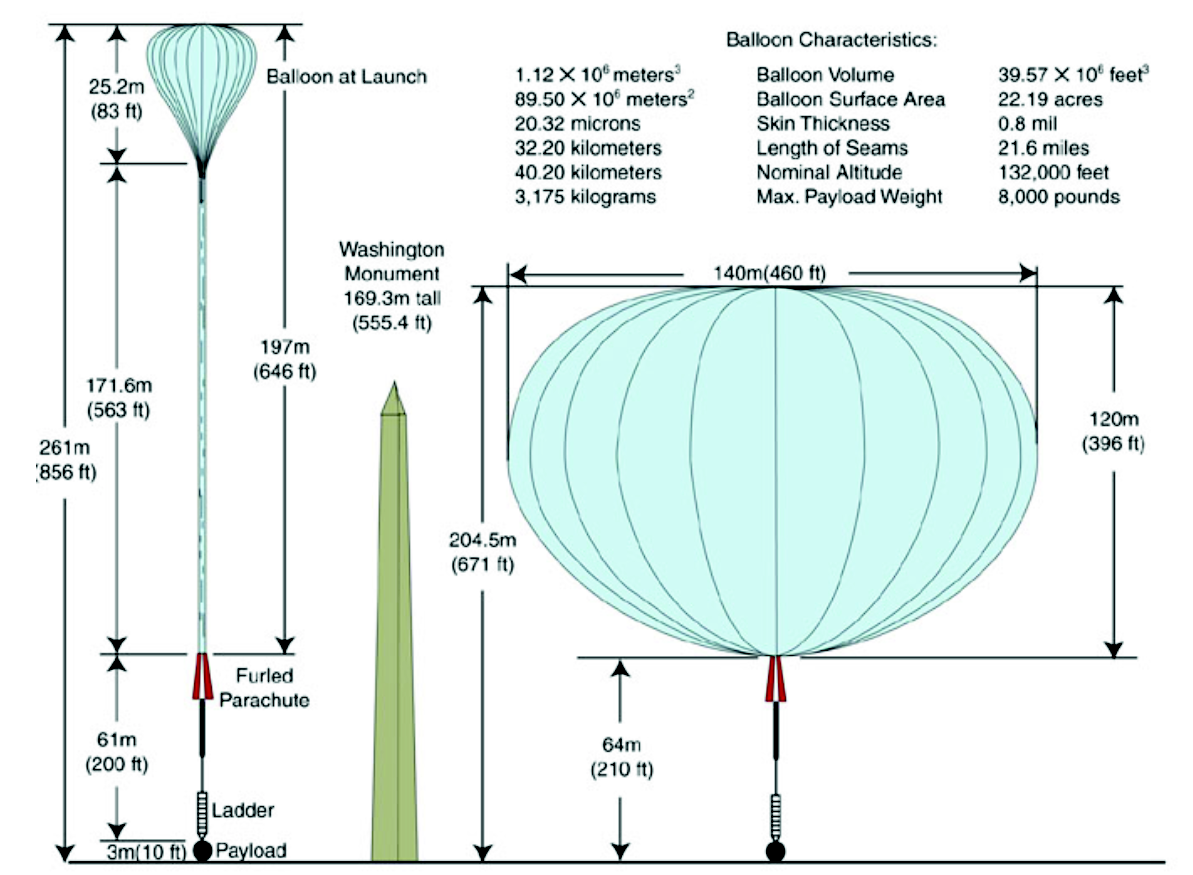}
\caption{A diagram showing the scale of balloons and the balloon train.}
\label{balloonscale}
\end{center}
\end{figure*}

Balloon-borne payloads are active contributors to all of these particle astrophysics topics.
In the following report, we provide a ``user's view'' of the current state-of-the-art of
NASA ballooning and highlight a number of the current and planned missions. While this will
not be a complete review of the NASA particle astrophysics program, we hope it will provide
a useful introduction to an exciting and accessible part of the NASA astrophysics program, 
one that is often overshadowed by spacecraft missions, but has produced a very
compelling portfolio of scientific accomplishments in the last few decades.

\section{Overview of balloons \& capability}

Before turning to the payloads themselves, we first summarize the
payload and flight profile capabilities of the program.

\begin{figure*}[htb!]
\begin{center}
\includegraphics[width=1.0\textwidth]{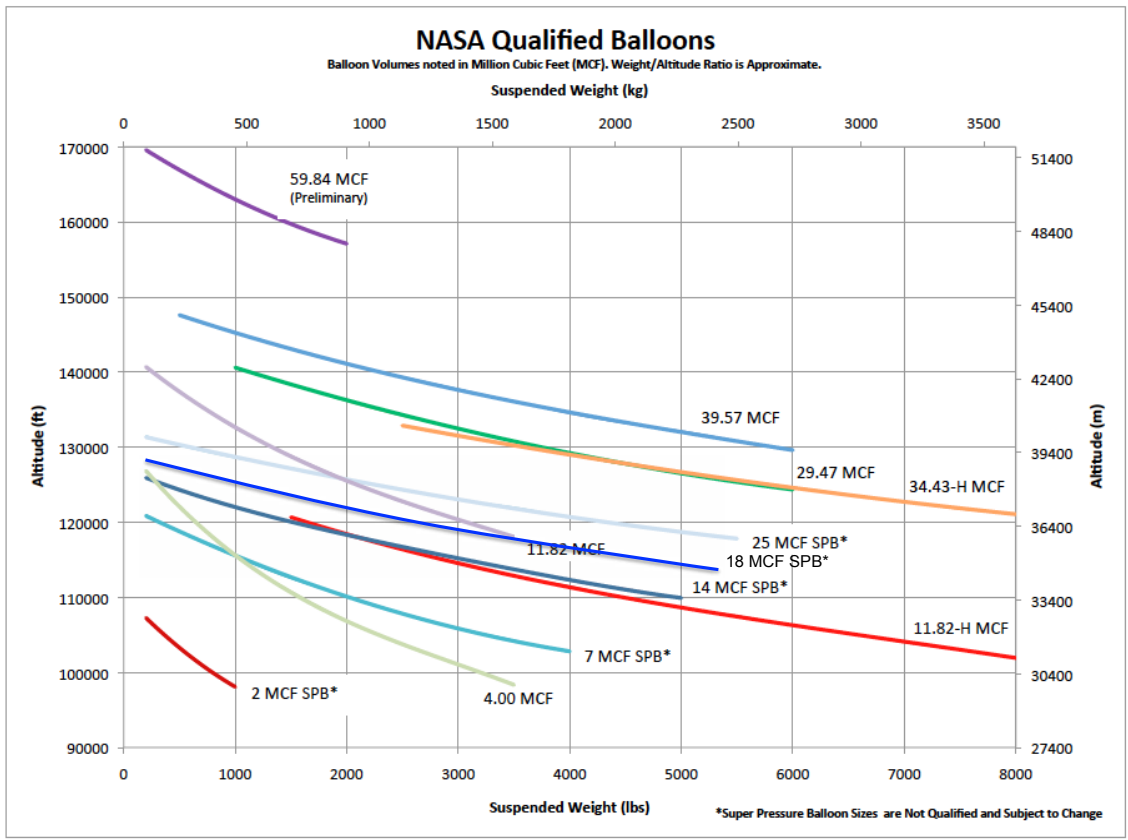}
\caption{Graphical summary of lift capacity for NASA qualified balloons.
Here the units MCF is the balloon volume in millions of cubic feet (there are
about 35.3 cubic ft per m$^3$). }
\label{qual}
\end{center}
\end{figure*}

Fig.~\ref{balloonscale} shows a graphical summary of the scale of a typical balloon in various 
stages of the launch and ascent to float. Helium initially inflates only a relatively small
volume of the balloon at ground level, and the flight train is typically around 800 ft long
at liftoff. Balloon launches and flights produce very low dynamics compared to a
rocket launch, and thus payloads may dispense with the rigorous vibration qualification
required of spacecraft payloads. The only significant dynamic event is at termination after the
flight is complete, and the parachute opens after several tens of km free fall from the float
altitude (typically 120Kft) down to about 80Kft, where the atmosphere has enough density to
fill the parachute. By this time the payload is falling at a high velocity, and the 
jerk from the filling chute can produce several gees of acceleration. Recent modifications
to the flight train using dynamic load absorbers have greatly reduced even these dynamics,
but payloads are still required to be designed for a maximum 10 gee acceleration at this
stage. However, the requirement is only that the payload does not separate from the
flight train, and this takes place after the science mission is complete.Thus the science
instrument does not have to be designed to operate after this acceleration event, only
to survive it in a safe-mode state.

In Fig.~\ref{qual} we show current altitude vs. lift capabilities for NASA qualified balloons. 
These come in both zero-pressure balloons (ZPB), and recently now in super-pressure (SPB) versions. 
ZPBs are vented and stay at equilibrium density/pressure with the surrounding gas at float.
Diurnal variations of solar heating and night-time cooling lead to substantial changes in
the internal pressure of the gas, and the equilibrium conditions thus cause a diurnal change
in altitude and balloon envelope shape for a ZPB. SPBs were developed to address this
drawback in ZPB float profiles, as we discuss below.

The largest ZPBs now routinely carry 2000 kg or more of science weights to altitudes over 
130,000 ft (nearly 40km), where the atmospheric overburden is below 3 mbar, or less than
0.3\% of sealevel pressure. Flights in Antarctica have now exceeded 55 days at
float, and 1-month-at-float has become a more routine expectation as NASA and the National
Science Foundation, which manages the Antarctic program, have developed more effective
protocols for log-duration flight support. 

While costs for such missions have grown steadily as
more complex instruments are developed and required for more challenging science, 
typical missions are still at least an order of magnitude and often two orders of
magnitude smaller than equivalent costs to get a comparable payload to low-earth orbit.
Thus it has become common to use balloon payloads to validate instruments prior to
their promotion as a potential spacecraft instrument, and this has the added benefit
of introducing new investigators to the NASA flight hardware development process
under a much lower risk and lower cost environment.

\subsection{Superpressure balloons.}

\begin{figure}[htb!]
\begin{center}
\includegraphics[width=0.5\textwidth]{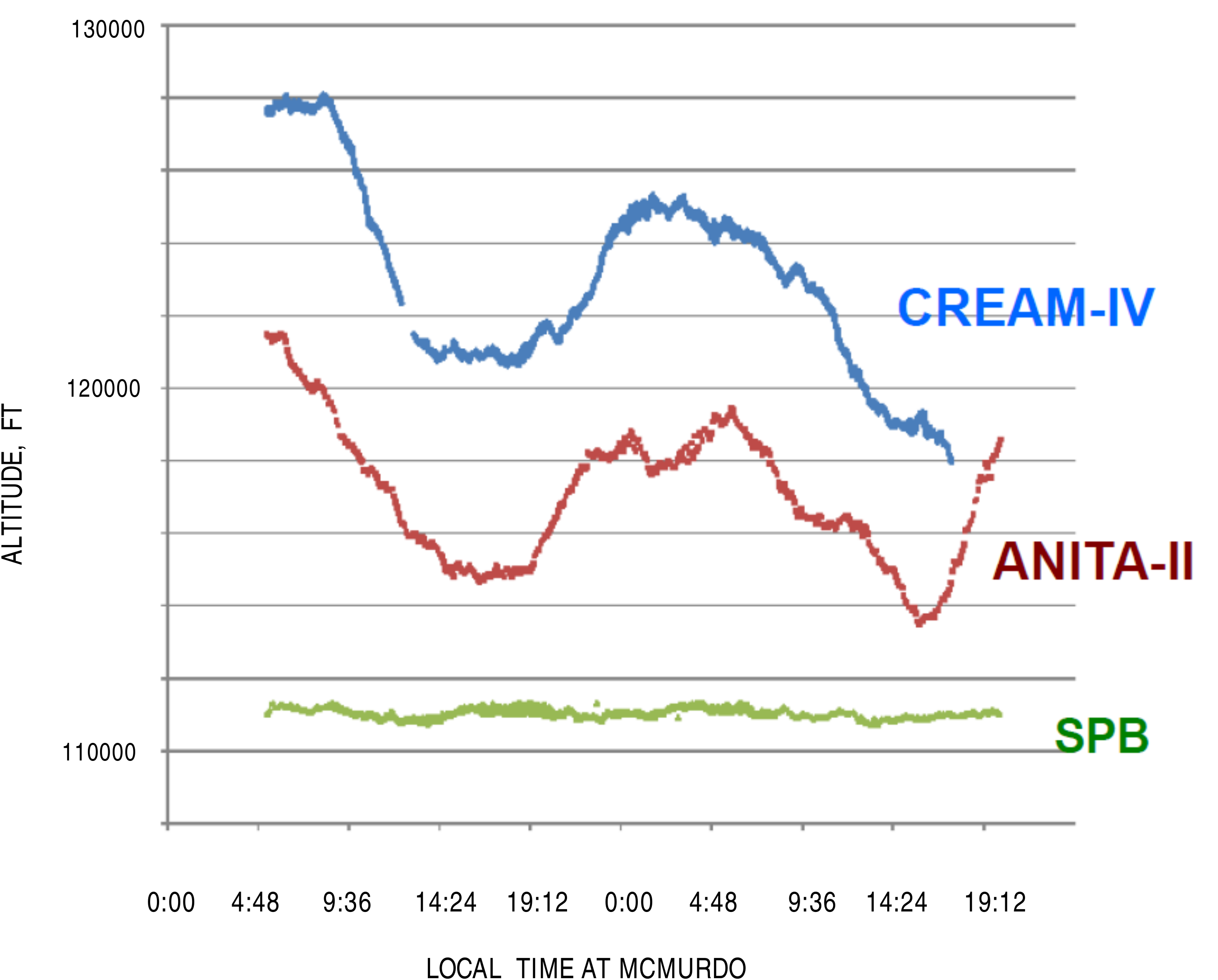}
\caption{Example of vented zero-pressure balloon altitude profiles for three
payloads aloft during the 2008-2009 Antarctic season, showing the nearly
constant altitude of the super-pressure payload compared to the 
diurnal behavior of the zero-pressure payloads.}
\label{spb_v_vented}
\end{center}
\end{figure}

In contrast to ZPBs, SPBs are not vented, and are designed to maintain a slight
overpressure (typically less than 100 Pa -- compare this to sea-level atmospheric pressure of
$10^5$~Pa) relative to the ambient atmosphere at float. This leads to a vast improvement in their
float altitude profile, and almost completely suppresses the diurnal altitude changes.
This is illustrated by Fig.~\ref{spb_v_vented} which shows the altitude profiles for
three payloads that were aloft at the same time during the 2009 austral summer season
in Antarctica: ANITA, CREAM, and a 7Mcft SPB test flight, which flew for 54 days during
that year. While ANITA and CREAM saw diurnal variations of 8-10Kft over the 1.8 day
period shown here, the variations for the SPB were in the tens of m RMS range.

\begin{figure}[htb!]
\begin{center}
\includegraphics[width=0.5\textwidth]{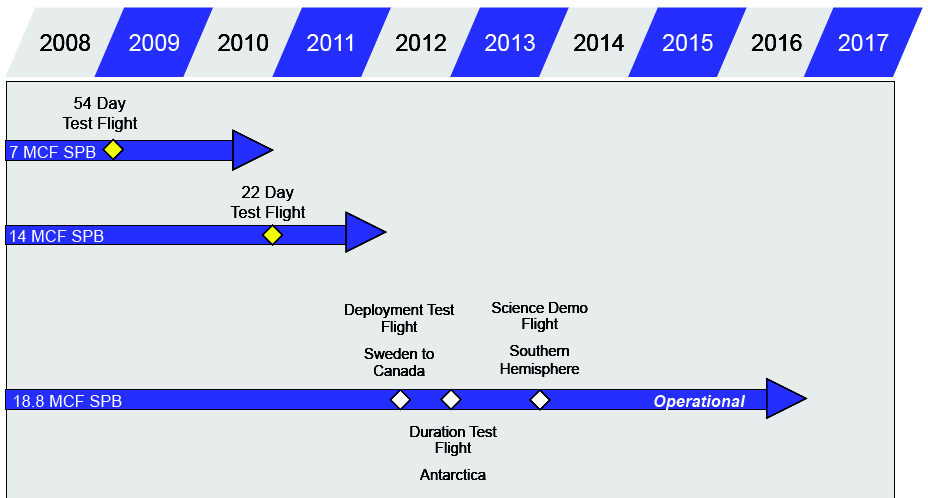}
\caption{Super-pressure balloon development schedule.}
\label{SPBsched}
\end{center}
\end{figure}

Super-pressure balloons are still largely in a development phase within the NASA
balloon program, although they are now considered part of the proposable launch
vehicle inventory, up to the 18 MCF size, although this latter balloon is still in
its final testing phases prior to being available for production. The current schedule
for the SPB program is shown in Fig.~\ref{SPBsched}, including the proof-flights for
the 7MCF and 14MCF balloons, both of which are capable of carrying quite substantial
payloads, a thousand kg or more to at least 110Kft.

One of the most important roles that SPBs will play in the future of the balloon program
is that they will create the possibility of mid-latitude flights for up to 100 days,
engendering payloads that wish to make night-side observations in these regions without
paying the stiff penalty of a loss of altitude during the night-time cooling period.
To facilitate such flights, southern hemisphere launch sites must be identified which will
allow for a complete orbit around the Earth, with overflight agreements from the nations
in the flight path. Currently the best potential site appears to be in Argentina,
but as yet no agreements have been reached regarding the overflight requirements.

\section{Antarctic flights.}

\begin{figure*}[htb!]
\begin{center}
\includegraphics[width=0.8\textwidth]{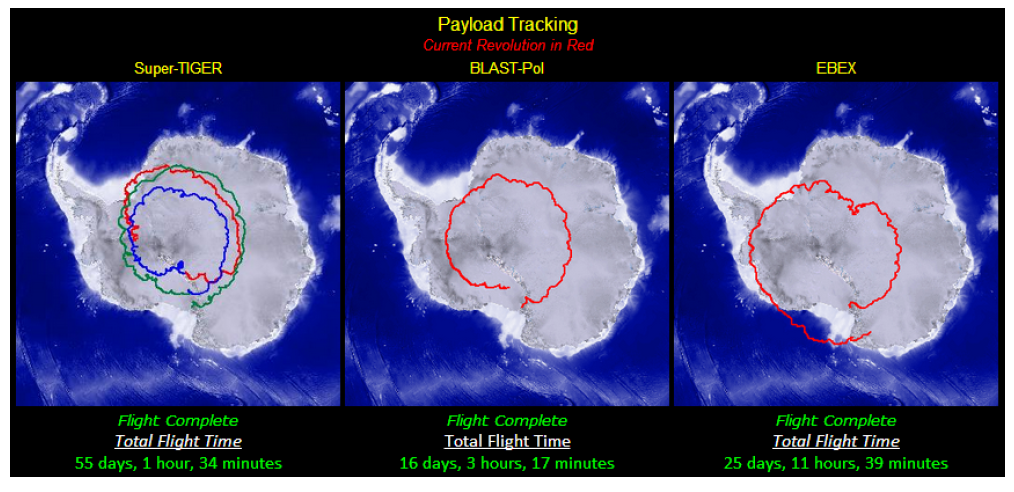}
\caption{Flight paths for payloads in the most recent Antarctic season,
including the record-breaking Super-TIGER flight of 55 days.}
\label{tracks2012}
\end{center}
\end{figure*}

Lacking a mid-latitude launch sight and corresponding overflight plan, Antarctica 
remains the centerpiece of scientific ballooning, affording the longest possible flights.
These are made during the austral summer months, early December through early February,
when the Polar Vortex creates stable circumpolar flight paths.
Payloads can thus also be assured of continuous solar illumination, allowing them to derive
all of their operational power from photovoltaics. The drawback is of course that no
night flights are yet possible; launch operations during austral winter are not yet
under consideration, and in any case circumpolar orbits would not be possible since the
stratospheric wind patterns are completely different than the summer months.

The very long possible exposures at float have created a demand for these flight 
opportunities, and NASA and NSF have responded by expanding the program recently
to allow for up to three large payloads to be launched in each of the Antarctic seasons.
Fig.~\ref{tracks2012} shows the results from the most recent season, where three payloads,
Super-Trans-Iron Galactic Element Recorder (Super-TIGER), Balloon-borne
Large-Aperture Sub-millimeter Telescope (BLAST-Pol), 
and E and B Experiment (EBEX, a cosmic background telescope) all completed successful missions, with Super-TIGER
breaking the record for the longest flight to date, 55 days over three orbits of the pole.
This latter flight illustrates nicely the behavior of the south Polar vortex, which is
characterized by stable stratospheric circulation. 

Launches are made from near
McMurdo Station, at Williams Field, about 10 miles from McMurdo on the Ross ice shelf,
which is about 50 meters thick at the launch site, floating over the Ross sea.
The shelf is gradually moving toward Ross island, at a rate such that the base has
been moved twice already over the last three decades since the program began. 
Despite this additional challenge, the smooth, flat surface of the ice shelf provides
and ideal platform for balloon launches, and the NASA program has become highly adapted
to this location, with two large payload integration buildings, and a host of other
supporting facilities, and (as many have noted) and a galley with 
some of the best food in Antarctica!

\begin{figure}[htb!]
\begin{center}
\includegraphics[width=0.35\textwidth]{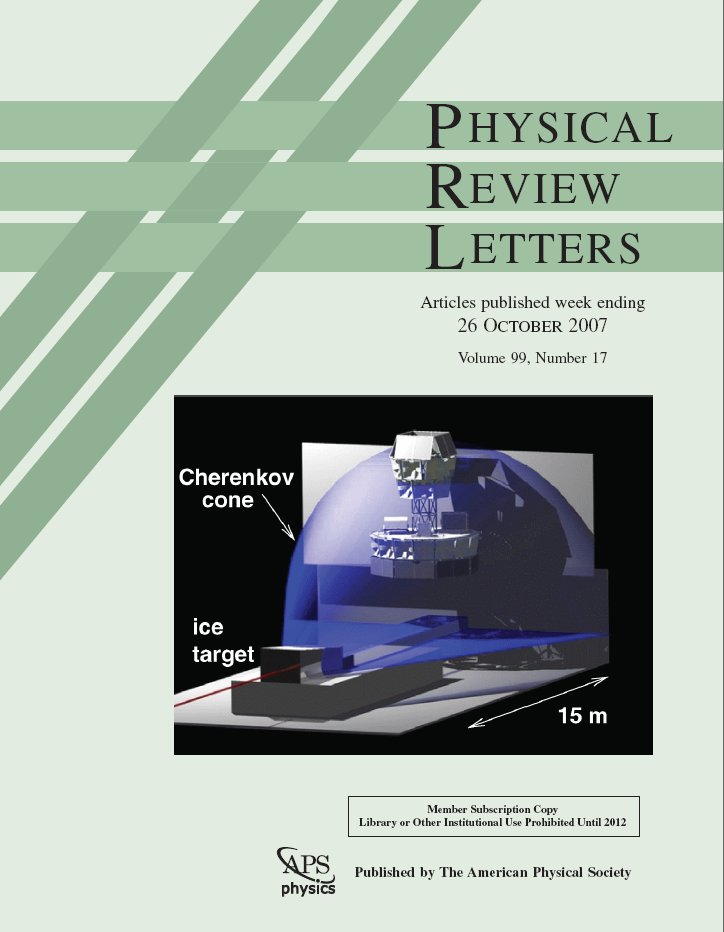}
\caption{PRL cover for ANITA measurement of the Askaryan effect in ice.}
\label{PRL1}
\end{center}
\end{figure}

\begin{figure}[htb!]
\begin{center}
\includegraphics[width=0.35\textwidth]{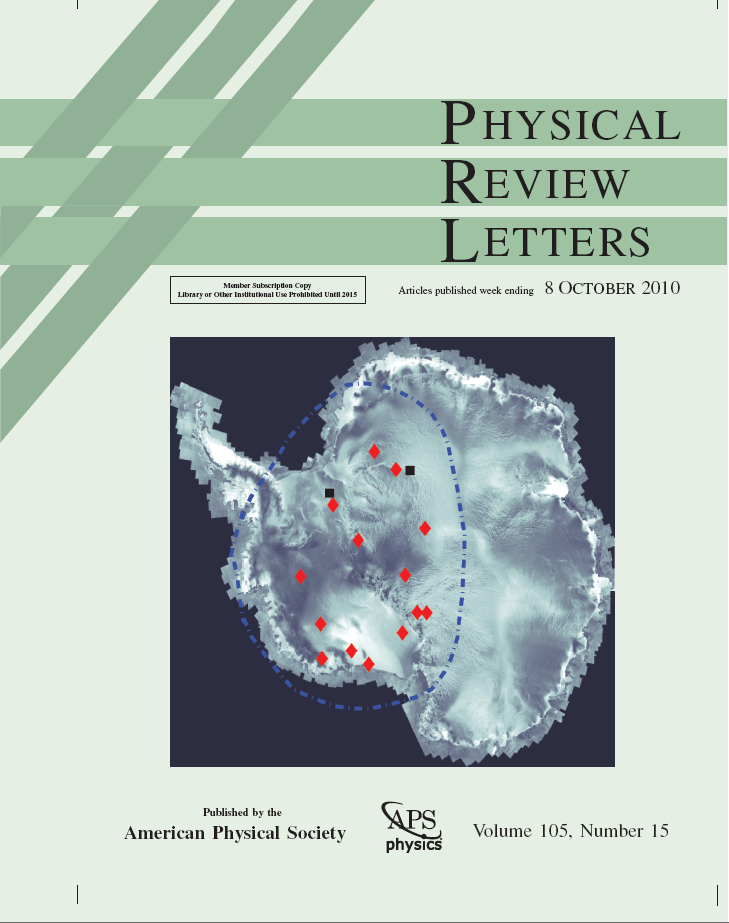}
\caption{PRL cover for ANITA-1 discovery of UHE cosmic rays.}
\label{PRL2}
\end{center}
\end{figure}

Since the advent of Antarctic long-duration ballooning in the early 1990's,
there have been several notable scientific results that bear repeating:
\begin{itemize}

 \item Boomerang flew in the early 1990's, creating one of the first detailed maps
of Cosmic Microwave Background (CMB) temperature fluctuations, and
demonstrating the Euclidean geometry of the universe.  This
led to the 2006 Balzan Prize for Astronomy and Physics~\cite{Boomerang},
and was instrumental in engendering follow-on space missions such as the
Cosmic Background Explorer (COBE) and its successors WMAP and Planck.

\item In a 2007 flight the Advanced
Thin Ionization Calorimeter (ATIC) measured an unexpected excess of 300 GeV cosmic ray
electrons~\cite{ATIC}. These observations, combined with positron evidence from
the PAMELA spacecraft, led to speculation that a dark matter particle annihilation
signature might be the origin of the anomaly, although current belief now favors
a relatively nearby unidentified object such as a pulsar, as the source.

\item The 2006-2007 flight of the Trans-Iron Galactic Element Recorder (TIGER) 
was used to make precise measurements of the abundances of elements from 26Fe to 34Se,
leading to some of the best evidence that acceleration of galactic cosmic rays 
takes place in OB star associations~\cite{TIGER}. The record-setting 2012-2013 Super-TIGER flight
should provide much higher precision measurements of refractory element abundances.

\item The Balloon
Experiment with a Superconducting Spectrometer (BESS), a joint Japanese-American
collaboration, flew nine times since 1993, and has made  measurements
of low-energy antiprotons and searches for anti-helium at levels of precision
that lead to useful constraints on certain dark matter models, especially in
combination with other experiments such as ATIC and the space-based platforms~\cite{BESS}.

\item  The Cosmic Ray Energetics and Mass (CREAM) experiment has reported a
hardening of the cosmic ray energy spectrum below the knee, and this feature 
appears to be in tension with standard models of propagation for galactic cosmic rays~\cite{CREAM}
This result was enabled largely by the six total flights that CREAM was able to
achieve in the program, leading to about 160 days of exposure livetime, the largest
of any balloon-borne experiment to date.

\item The Antarctic Impulsive Transient Antenna (ANITA) mission was flown initially
in 2006-2007 as the first NASA ultra-high energy neutrino payload. In the process of
calibrating the payload at the Stanford Linear Accelerator Center (SLAC), ANITA
made the first measurements of the Askaryan effect -- radio Cherenkov emission from
the negative charge excess in an electromagnetic cascade -- in ice, using
a 7.5 ton block-ice target in the SLAC End Station A experiment hall. This
experiment was chose for the cover of Physical Review Letters, the premier fundamental
physics journal worldwide, as seen in Fig.~\ref{PRL1}. 

\item Then,
during the first ANITA flight, the payload not only set the best world limits on
UHE neutrino fluxes~\cite{ANITA1,ANITA2}, but made a serendipitous observation of radio emission from
16 ultra-high energy cosmic rays, which interacted in the Antarctic atmosphere, 
producing a narrow radio beam which then reflected off the ice surface up to
the payload. This unexpected discovery was not the first time radio pulses were
seen from cosmic rays, but it was the highest energy set of such events ever
recorded, and was unique in that the instrument was completely self-triggered.
These led to a second Physical Review Letters cover, shown in Fig.~\ref{PRL2}~\cite{ANITA-CR}.

\end{itemize}

\section{Future Particle astrophysics payloads.}

As noted above the current very long near-space exposures possible for payloads
in the Antarctic long-duration balloon program has led to renewed interest in
these opportunities, and several new particle astrophysics efforts will be highlighted here.

\subsection{GAPS}

The General AntiParticle Spectrometer experiment (GAPS) instrument is designed to
search for low-energy cosmic-ray antideuterons. The detection of any such events would
be significant, since although antideuterons are produced in interactions in the interstellar
medium, the kinetic energies from those processes are expected to be high compared to
the kinetic energy of antideuterons generated by dark matter annihilation. Thus low-energy
antideuterons provide a dark matter detection channel, if sufficient sensitivity and particle
identification can be achieved. GAPS has developed a novel approach using 
a Lithium drifted silicon tracker coupled to a time-of-flight system to separate antiprotons from
antideuterons, and the approach promises to extend the low-energy reach below any other existing
instrument if it is successful. 
GAPS has completed a prototype balloon flight which validated the basic system, but much work
is still needed to achieve the low-energy sensitivity needed~\cite{GAPS}.

\subsection{EVA}

\begin{figure}[htb!]
\begin{center}
\includegraphics[width=0.5\textwidth]{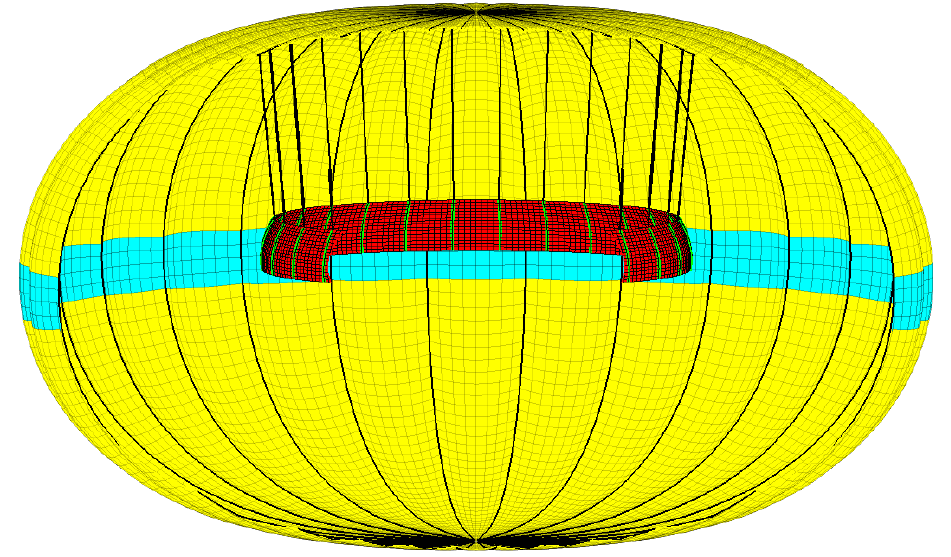}
\caption{A finite element model of the EVA balloon, including the reflective band around the equator (in light blue)
and the feed antenna support membrane (red) in the balloon interior. This model has a small number of exaggerated
gores; the full-scale model has about 280 gores, but their relief is relatively low and thus the optics are not
degraded at radio wavelengths.}
\label{EVA-cut}
\end{center}
\end{figure}

The ExaVolt Antenna (EVA) mission is currently funded for NASA technology development toward
a mission which would center on the use of a super-pressure balloon. EVA's goal is the detection
of ultra-high energy neutrinos and cosmic rays through observations over Antarctica,
following in the spirit of the successful ANITA payload series, but with a plan to increase
the sensitivity by as much as two orders of magnitude~\cite{EVApaper}.

EVA's novel approach makes use of toroidal radio reflector/concentrator optics that are
integral to the superpressure balloon itself. A $\sim 10-12$~m high reflective equatorial
band on the outer balloon membrane forms a powered toroidal radio-frequency mirror, with a focus inside the
balloon, about halfway between the center and the portion that is reflecting. The optics
are such that the balloon can receive radio impulses from any direction near the horizon,
covering a very large synoptically viewed area of 1-2M square kilometers. At the focal
zone inside the balloon, a membrane is suspended containing a feed antenna array which
receives the power from the focusing optics. Fig.~\ref{EVA-cut} shows 
a cutaway view of the basic geometry of the system. The light blue band represents the reflector band,
and the red band in the interior is the feed antenna support. This finite element model had only a small
number of gores; in practice the 18MCF full-scale version will have 280 gores. 

EVA's optics must be stable and smooth enough to avoid degrading reflected RF signals up to frequencies
of several hundred MHz; this is only possible using a SPB, since the overpressure maintains a stable
surface profile. Physical optics simulations and exact microwave scale models have confirmed that
the optics perform well at the required radio wavelengths, and a 1/10 to 1/5-scale demonstration balloon
is expected to be fabricated sometime in 2014. If the approach is validated and the technology
readiness level can be raised, EVA can be expected to be a contender for a near-future Small Explorer
Mission of Opportunity.

\subsection{Pathways to (and from) Space instruments.}

The success of the CREAM payload has now led to selection of a CREAM follow-on for the 
International Space Station (ISS), given the acronym ISS-CREAM (pronounced like the ice-cold dessert).
Because of the extended mission now given the ISS, engendered by 
the continuing support for the Alpha Magnetic Spectrometer, ISS-CREAM will allow an upgraded version of
the CREAM instrument to gather much higher statistics needed to resolve the behavior of the cosmic ray
spectrum at the knee. ISS-CREAM is expected to be launched in 2014.

\begin{figure}[htb!]
\begin{center}
\includegraphics[width=0.5\textwidth]{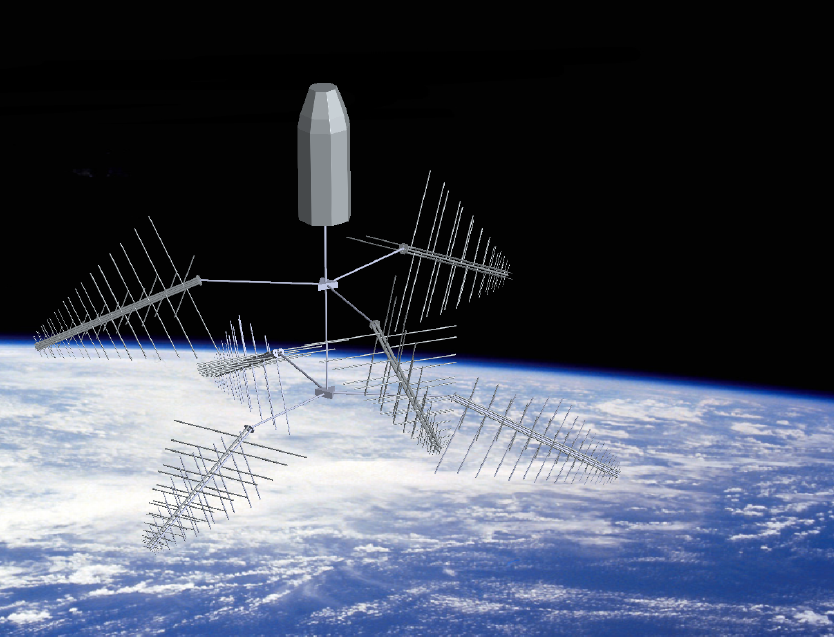}
\caption{Artist conception of the SWORD spacecraft, with its antenna array fully deployed.}
\label{sword}
\end{center}
\end{figure}

In a twist, a mission planned for many years for the ISS, the Japanese Experiment Module Extreme Universe
Space Observatory (JEM-EUSO), which is slated to be lifted to the ISS late this decade, will first
validate some of its required measurements using a balloon payload. JEM-EUSO uses a very large
Fresnel-lens and associated fast camera to image the optical fluorescence emission from UHE cosmic ray
air showers transiting the atmosphere below it, from an altitude of 350 km on the ISS. By careful timing
of the arrival of the fluorescence pulse, JEM-EUSO can determine the arrival direction of the primary
particles, and by integrating the received light, establish their energy. JEM-EUSO will be the first
UHE cosmic ray observatory in space, but to ensure that the measurements can be made with the
requisite accuracy, the EUSO-Balloon mission will attempt three flights in the 2014-2016 time frame
with an instrument which will allow an end-to-end test of JEM-EUSO's key technologies.

Finally we note that ANITA's discovery that a balloon payload could successfully observe ultra-high
energy cosmic rays has generated interest in extending this method of observation to a space-based
platform, where the acceptance could be large enough to greatly extend the reach compared to
even the largest ground-based observatories, providing a method that is highly complementary to
the approach used by JEM-EUSO.

The Synoptic Wideband Orbiting Radio Detector (SWORD) 
satellite mission, now under development at NASA's Jet Propulsion Lab, is a new concept
based on ANITA's results~\cite{SWORD}, and shown in concept in Fig.~\ref{sword}. 
SWORD will fly a deployable interferometric antenna array
covering the 30-300 MHz range, lower than that used by ANITA, but more tuned to the largest
amplitude radio emission from UHE cosmic rays. SWORD observes UHECRs through the reflection of
the impulsive beamed radio emission off the Earth's surface, whether ocean, land, or ice. Challenges
faced by SWORD will be calibration of the reflective surface for each event, along with deconvolution
of the ionospheric effects, which are quite strong, especially at the low end of SWORD's frequency range.

SWORD is planned as a low-cost SMEX mission, but could achieve sensitivity of nearly two orders
of magnitude compared to ground-based observatories, particularly in the super-GZK energy range,
above $10^{20}$~eV, where little is known about the endpoint behavior, and the sources should become
more evident due to the much lower deflection by intergalactic magnetic fields. SWORD is planned
to fly at the next solar minimum, roughly 2019.

In the interim, ANITA will fly again in 2014, and should measure several hundred more radio-detected
UHECRs. This will go a long way toward refining the radio emission models that are necessary
to more precisely calibrate the response of instruments like SWORD, and it is likely that an
ANITA-IV instrument will be proposed for the Antarctic LDB program, with a specific goal of measuring
even more such events. 

\section*{Acknowledgments}
This work was supported by the  NASA balloon program, the US Department of Energy, Office of Science,
through the High Energy Physics Program, and the National Science Foundation through
their Antarctic Program. We also thank the Columbia Scientific Balloon Facility
for their excellent work in supporting NASA's balloon program.




\end{document}